# Title: The Missing Covariate Indicator Method is Nearly Valid Almost Always

Gang Xu†, Mingyang Song†, Xin Zhou, Yilun Wu, Mathew Pazaris, and Donna Spiegelman*

***Corresponding author: Donna Spiegelman**, Department of Biostatistics, Yale School of Public Health, New Haven, Connecticut, USA, Email: donna.spiegelman@yale.edu
**Gang Xu**, Department of Biostatistics, Yale School of Public Health, New Haven, Connecticut, USA; Laboratory of Neurogenetics and Precision Medicine, University of Nevada, Las Vegas, Nevada, USA
**Mingyang Song**, Departments of Epidemiology, Harvard T.H. Chan School of Public Health, Boston, Massachusetts, USA
**Xin Zhou and Yilun Wu**, Department of Biostatistics, Yale School of Public Health, New Haven, Connecticut, USA
**Mathew Pazaris**, Fidelity Investments, Boston, Massachusetts

†: **Gang Xu** and **Mingyang Song** contributed equally to this study.

**Abbreviations:** HPFS, Health Professionals Follow-up Study; MAR, missing at random; MCAR, missing completely at random; MCIM, missing covariate indicator method; MI, multiple imputation; CC, complete case; NM, no missingness; ARE, asymptotic relative efficiency; NHS, Nurses' Health Study; RR, relative risk; OR, odds ratio.

**Running Head:** Missing Covariate Indicator Method

# Abstract

Although the missing covariate indicator method (MCIM) has been shown to be biased under extreme conditions, the degree and determinants of bias have not been formally assessed. We derived the formula for the relative bias in the MCIM and systematically investigated conditions under which bias arises. We found that the extent of bias is independent of both the disease rate and the exposure-outcome association, but it is a function of 5 parameters: exposure and covariate prevalences, covariate missingness proportion, and associations of covariate with exposure and outcome. The MCIM was unbiased when the missing covariate is a risk factor for the outcome but not a confounder. The average median relative bias was zero across each of the parameters over a wide range of values considered. Our simulation study demonstrated that the mean and median of relative bias of MCIM was comparable to that of the no missingness method, which used the full sample with complete information for all variables, as long as the missingness of covariate is independent of the outcome. When missingness was no greater than 50%, less than 5% of the scenarios considered had relative bias greater than 10%. In several analyses of the Harvard cohort studies, the MCIM produced materially the same results as the multiple imputation method. In conclusion, the MCIM is nearly valid almost always in settings typically encountered in epidemiology and its continued use is recommended, unless the covariate is missing in an extreme proportion or acts as a strong confounder.

# 1 Introduction

Missing data is a common problem encountered in many epidemiologic and clinical studies. It can occur in any of the variables in a study, including exposure, outcome, or covariates that may or may not be confounders of the exposure-outcome relationship [1]. In epidemiology, we typically treat missing data in potential confounders differently than missing data in the primary outcome and exposure variables for analysis. For example, missing data on outcomes can seriously compromise inferences from clinical trials, and robust prevention and treatment measures have been summarized [2]. In observational studies, participants without data on the primary exposure or outcome are typically excluded from the study. That is, not having data on the outcome of interest or the exposure under investigation is a primary exclusion criterion for most studies. If participants who are included in the study population are different in ways that lead to bias in the estimated measure of exposure-outcome association, selection bias results. In general, selection bias in the relative risk (RR) will result when the probability of being included in the study population depends jointly on exposure and outcome status, after properly controlling for confounders [3]. Methods to deal with selection bias have been an active research area and will not be discussed any further here [4]. In this study, we focus on missingness in covariates.

As the simplest approach, complete-case (CC) analysis is inefficient and can be biased, because subjects with complete data recorded for all covariates can be a small and biased subsample of the study subjects [5]. Other more sophisticated modeling or imputation-based approaches have been developed during the past few decades, such as inverse probability weighting, multiple imputation (MI), and maximum likelihood [6]. Although these methods are theoretically appealing, their validity is dependent on correct specification of an additional model [7], invoking what are often empirically unverifiable assumptions that can pose even greater challenges for complex models in longitudinal settings [8]. An additional practical limitation is their computational cost that may be prohibitive in large epidemiologic studies and even more so in this era of ‘big data’.

Another simple approach for missing covariates is the indicator method (MCIM), where a missing indicator is created and added to the model for each variable with missing data [9]. The missing indicator takes value 1 whenever the original variable is missing, and 0 otherwise. We then assign the value of 0 to the original covariate for all those originally missing on the covariate. Formally, let $M_i$ be the missing indicator for participant $i$ and $C_i$ be participant $i$'s value of the covariate that is sometimes missing. The new variables are $M_i$ and $(1 - M_i)C_i$. To apply the MCIM method to a stratified 2x2 table analysis, we stratify by both the missing indicator and the recoded original variable (with missing values replaced by 0) to estimate the relative risk, or, in a regression analysis, we control for both $M_i$ and $(1 - M_i)C_i$ in the model.

Because all participants are included in the analysis, the missing indicator method is more efficient than the complete-case analysis. Although the missing indicator method has been shown to be biased under extreme conditions [10, 11], it is widely used for missing covariates [12] and has been recommended as a missing data method for propensity score analysis [13, 14]. However, the degree and determinants of bias have not been formally assessed, nor have the conditions under which it is unbiased.

In this study, we conducted a numerical study of the bias arising from the MCIM and conducted a simulation study comparing the MCIM to no missingness (NM) method, MI method, and CC method. The NM method utilizes all samples with complete information for all variables in the simulation study, whereas CC method uses the subset of the data with no missing values. To put our findings in the context of large cohort studies, we also compared the results from previously published studies that have used both the MCIM and the multiple imputation method.

# 2 Methods

## 2.1 Evaluation of bias.

We considered a binary exposure ($E$), outcome ($Y$), and covariate ($C$). For the covariate $C$ that is missing in some participants, the missing indicator method stratifies the data by three levels of the covariate: $C = 1, 0$, and missing (**Table 1**). For simplicity, we first assume that the missing covariate mechanism is completely at random (MCAR), that is, that the probability of missingness of the covariate is independent of any observed or unobserved data included in the primary outcome model. The stratified 2x2 tables for $C = 1$ and $0$ are unconfounded by the covariate $C$ and thus give valid estimates for the effect of exposure on outcome. The stratum with missing covariate is a crude table, collapsed over the two levels of the covariate. Therefore, it may yield a biased exposure estimate if $C$ is indeed a confounder. The convergent value of the final summary effect estimate of relative risk, $RR_e$, is then a weighted average of the convergent value of one biased estimate, $RR_{miss}$, along with the convergent value of two valid estimates, $RR_{C=1}$ and $RR_{C=0}$. Assuming homogeneity of relative risks across strata formed by levels of the covariate, $C$, $RR_{C=1}$ and $RR_{C=0}$ are equal, and denoted by $RR(E)$. It is usually the case that this assumption is valid [15].

**[Place Table 1 near here]**

Thus, $RR_e = [1 - \Pr(C_{miss})] * RR(E) + \Pr(C_{miss}) * RR_{miss}$, where $\Pr(C_{miss})$ is the proportion of participants with the missing covariate. We derived an expression for $P_{bias}\%$ as a function of 5 underlying parameters (see Appendix 1): the prevalence of the exposure, $Pr(E)$, the prevalence of the covariate, $Pr(C)$, the proportion of missingness, $Pr(C_{miss})$, the effect of the covariate on the outcome, $RR(C)$, and the association between the exposure and the covariate, $RR(E|C)$.

$$P_{bias}\% = \Pr(C_{miss})\left(\frac{\Pr(C)\,[1-\Pr(C)]\,[RR(C)-1][RR(E|C)-1]}{[1-\Pr(E)][1-\Pr(C)+\Pr(C)\,RR(C)RR(E|C)]-\Pr(C)\,[1-\Pr(C)][RR(C)-1][RR(E|C)-1]}\right) \times 100\%. \quad (1)$$

In many epidemiologic and clinical studies, the parameter of interest is the odds ratio. If the prevalence of the outcome, that is, $Pr(Y)$, is low, which is often the case, the odds ratio approximates the relative risk well. Then, formula (1) applies to the odds ratio as well.

In Appendix 2 and 3, we investigated the conditions under which the above formula (1) remains valid when the missingness mechanism of covariate data is missing at random (MAR), that is, when the probability of covariate missingness depends only upon variables that are never missing in the data. Under MCAR, the observed exposure relative risk for each covariate stratum ($C = 0$ or $C = 1$) converge to the true relative risk $RR(E)$ and the crude RR in the missing covariate stratum equal to the RR in the complete data stratum. However, under MAR, these conditions may not hold, thereby invalidating the formula (1).

In appendix 2, we demonstrated that formula (1) is valid when missingness of covariate $C$ is independent of the outcome $Y$, even if it depends on the exposure $E$. Under this condition, the observed exposure RR for each covariate stratum converges to the true relative risk $RR(E)$ and the crude RR in the missing covariate stratum equals to the RR in the complete data stratum.

In Appendix 3, we extended this investigation to rare diseases, where the odds ratio (OR) is a close approximation of the risk ratio (RR). Because many chronic diseases are rare, the OR approximates the RR well. We showed that formula (1) remains approximately valid under MAR when the missingness of covariate $C$ is independent of either the outcome $Y$ or the exposure $E$. Under this condition, the observed exposure odds ratio for each covariate stratum converges to the true odds ratio $OR(E)$, and the crude odds ratio in the missing covariate stratum equals the odds ratio in the complete data stratum.

**2.2 Numerical Bias Evaluation.**

Based on the ranges typically encountered in epidemiologic studies, we assigned a range of values to each of the 5 determining parameters for $P_{bias}\%$ (**Table 2**). For example, we allowed the relative risk estimate for the effect of the covariate on the outcome to range from 1/5 to 5. We then calculated the $P_{bias}\%$, the quantity of interest in this study, based on all 33,540 valid combinations of the values considered for each of the 5 parameters over the range through a closed-form analytic expression for $P_{bias}\%$. Furthermore, based on formula (1), $P_{bias}\%$ does not depend on Pr $(Y)$ and $RR(E)$, so these two parameters were excluded from the numerical bias evaluation.

**[Place Table 2 near here]**

**2.3 Simulation Study to Evaluate Bias of MCIM.**

We conducted a simulation study to compare the bias of MCIM to three other methods, NM, MI, and CC. Additionally, we evaluated the asymptotic relative efficiency (ARE) of MCIM, MI, and CC with respect to NM. Our study considered two missing data mechanisms: MCAR and MAR. For MAR, we simulated the missing indicator variable through a logistic model:

$$logit\left(P(M_i = 1|E_i, Y_i)\right) = \alpha_0 + \alpha_1 E_i + \alpha_2 Y_i, \text{ for } i = 1, \cdots, 2{,}000,$$

where $\alpha_0$ is the intercept, $\alpha_1$ is the effect of $E_i$ to $M_i$, and $\alpha_2$ is the effect of $Y_i$ to $M_i$. We considered scenarios where $M$ is independent of $Y$ ($\alpha_2 = 0$) and where $M$ is related to $Y$ ($\alpha_2 = log(2)$).

Due to computational limitations, we did not explore all parameters combination for numeric bias evaluation. Instead, we used a subset of parameter combinations as shown in **Table 2**. In addition to five parameters, we varied four additional parameters, including $\Pr(Y)$, $RR(E)$, $\alpha_1$, and $\alpha_2$. Further details of the simulation settings are provided in Appendix 3.

To assess the exposure effect in the simulation study, we fitted a generalized linear model:

$$log\left(P(Y_i = 1|E_i, C_i)\right) = \beta_0 + \beta_1 E_i + \beta_2 C_i$$

for NM and CC, where $\beta_0$ is the intercept, $\beta_1$ is the exposure effect and parameter of interest, $\beta_2$ is the effect of covariate. For MCIM, we fitted the model:

$$log\left(P(Y_i = 1|E_i, C_i)\right) = \beta_0 + \beta_1 E_i + \beta_2 C_i(1 - M_i) + \beta_3 M_i,$$

where $\beta_3$ is the effect of missing indicator variable. Finally, for MI, we fitted the model:

$$log\left(P(Y_i = 1|E_i, C_i)\right) = \beta_0 + \beta_1 E_i + \beta_2 \hat{C}_i,$$

where $\hat{C}_i = C_i$, if $M_i = 0$, and $\hat{C}_i = \hat{C}_{MI,i}$, if $M_i = 1$, with $\hat{C}_{MI,i}$ being the imputed value of $C_i$ by MI when $C_i$ is missing. We used R package "jomo" to implement MI via Markov chain Monte Carlo under the assumption that $(C_i, E_i, Y_i)$ follows multivariate normal distribution [16] with the number of imputed datasets set to 10 in the simulation study. We excluded low-quality MI estimators if the outcome model converged with the imputed $\hat{C}_i$ less than 5 times, arose in approximately 31% of total runs.

**2.4 The extent of covariate missingness in large observational cohort studies.**

To examine the extent to which missingness occurs in some typical epidemiologic studies, we calculated the proportion of missing covariate data in two large cohort studies, the Nurses' Health Study (NHS) I and II, using breast cancer risk factors as an example. The NHS I is a prospective cohort which began in 1976 when 121,700 female nurses aged 30-55 years completed a mailed health questionnaire [17]. Similar questionnaires were returned in 1989 from the NHS II, comprised of 116,430 female nurses aged 25-42 years [18]. Follow-up questionnaires were mailed biennially to members of the two cohorts to update lifestyle and medical information. Diet was assessed using a validated food frequency questionnaire [19] every four years. We calculated the proportion of person-years for which the variables were missing among the total person-years. We considered a variety of reproductive and lifestyle breast cancer risk factors, including age at first birth, alcohol consumption, history of benign breast disease, family history breast cancer, age at menarche, age of menopause, physical activity, oral contraceptive use, postmenopausal hormone use, and body mass index.

**2.5 Comparison of the results produced by MCIM and MI.**

Because investigators who use MCIM in their primary analyses are occasionally asked by reviewers to use multiple imputation for missing covariate data, at least as a sensitivity analysis, we performed a head-to-head comparison of the results produced by the two methods in the five published studies [20-24], among thousands using data from three cohort studies, the NHS I and II and the Health Professionals Follow-up Study (HPFS), where this was requested. The HPFS is an observational cohort that enrolled 51,529 male health professionals aged 40-75 years in 1986. Similar follow-up procedures have been used as in the NHS I and II [25].

# 3 Results

## 3.1 When the covariate with missingness is a risk factor but not a confounder (or neither), the MCIM is unbiased

It is immediately evident from equation (1) that when $RR(C) = 1$ or $RR(E|C) = 1$, $P_{bias}\%$=0. Often, risk factors for the outcome are adjusted for in the analysis when they are not confounders; in many models, doing so will improve the precision of the estimate of the parameter of interest [26], here, $RR(E)$. Other times, to be conservative to strengthen the validity of a causal inference, investigators will adjust for known and suspected risk factors for the outcome even when they might not be confounders, in case they are. When this adjustment turns out to be unnecessary for validity purposes, as may often be the case, the MCIM method will be unbiased.

## 3.2 Almost no bias nearly always

Within each unique combination of the 5 parameters determining $P_{bias}\%$, we calculated the median, 25th and 75th percentile of $P_{bias}\%$, as well as the percentage of instances where $P_{bias}\%$ was higher than 5% and 10% (**Table 3**). The median of $P_{bias}\%$ was zero for each parameter value averaged over all the others, with the 25th and 75th percentiles below 0.5% in all but extreme cases, such as when the covariate missingness proportion exceeded 25%, or when the exposure or covariate was associated with a five-fold increase in risk of the outcome.

**[Place Table 3 near here]**

Furthermore, $P_{bias}\%$ exceeded 10% in only 1.1% of the parameter space explored. For example, $P_{bias}\%$ was greater than 10% in 4.8% of the scenarios considered when the covariate was missing in half of the study population, and was greater than 10% in 4.3% of the scenarios considered when the covariate was a strong confounder, with a relative risk for the outcome in relation to the confounder greater than 5. Even a $P_{bias}\%$ greater than 5%, as shown in the last column of **Table 3**, was a rare event in most scenarios considered here.

### 3.3 MCIM is unbiased when the missing indicator $M$ is independent of outcome $Y$

To assess the bias of exposure effect $\beta_1$, we calculated the relative bias of $\hat{\beta}_1$ as $\frac{\left(\bar{\hat{\beta}}_1-\beta_1\right)}{\beta_1}$, where $\bar{\hat{\beta}}_1$ is the average of $\hat{\beta}_1$ over the 2,000 simulated datasets. Firstly, we calculated the proportion of converged scenarios, defined as the number of scenarios where all four methods have converged estimators in at least one dataset over the total number of scenarios considered in the simulation study. We investigated the non-converged scenarios and identified separation (also known as perfect prediction) as the primary cause [27]. We then used these converged scenarios to calculate the mean and median percentage of relative bias for four methods for fair comparison. Additionally, we determined the proportion of scenarios with the absolute value of relative bias exceeding 5%, 10%, or 50%, over the total number of converged scenarios.

**Table 4** displays the results of proportion of converged scenarios and the relative bias for four methods. The MI method had the lowest proportion of converged scenarios compared to the other three methods. In MCAR and MAR with $M$ independent of $Y$, the NM, MCIM, and MI methods have similar means and median relative bias while CC has the largest mean and median of relative bias among all methods. In MAR with $M$ related to $Y$, the relative bias of MCIM was larger than NM and MI, but still smaller than CC. Furthermore, under MCAR or MAR with $M$ independent of $Y$, MCIM had the largest proportion of scenarios where the absolute value of relative bias exceeded $5\%$, $10\%$, or $50\%$ to other three methods.

**[Place Table 4 near here]**

We display the results of the ARE summary of MCIM, MI, and CC with respect to NM in **Table 5**. The ARE values for both MCIM and MI are close to 1, indicating similar asymptotic efficiency compared to NM. The CC's mean AREs were 0.85 under MCAR, 0.82 under MAR when $M$ is independent on $Y$, and 0.75 under MAR when $M$ is related to $Y$. These findings suggest that CC has lower asymptotic efficiency compared to NM.

**[Place Table 5 near here]**

### 3.4 Low missingness in large, well-established cohorts

We explored the extent of missing covariate data in some well-established cohort studies, using the NHS I and II as an example. We calculated the proportion of missing data among common risk factors for breast cancer measured in these studies. As shown in **Table 6**, most risk factors were missing less than $5\%$ of the person-time under follow-up and the extent of missingness rarely exceeded $10\%$. We see that although missingness is low overall in NHSII, even when missingness in any model covariate is taken into account ($7.4\%$). In NHS, if we were to use a complete case analysis instead of the missing indicator method, we would lose over $20\%$ of the data, likely unnecessarily, since, although being risk factors for the outcome, these variables are likely to be either weak confounders or not confounders at all. The standard practice in the Nurses' Health Studies is to control for them when we can, using the missing indicator method, providing at the very least, partial control.

**[Place Table 6 near here]**

### 3.5 No difference between results from MCIM and multiple imputation

Because MCIM has been considered a biased method, studies that use this method in their primary analysis are occasionally asked by journal reviewers to run additional analyses using more sophisticated methods for handling missing covariate data, such as multiple imputation. Therefore, to assess the extent to which the results are changed after switching from MCIM to multiple imputation, we compiled results from published studies in the three Harvard cohorts that have used both methods in the analysis. As shown in **Table 7**, MCIM yielded materially the same results as multiple imputation did in all cases considered.

**[Place Table 7 near here]**

## 4 Discussion

We derived an explicit expression for the bias associated with the use of MCIM. We were then able to show that when the covariate is not a confounder, MCIM is unbiased. Our simulation study demonstrated that the mean and median of relative bias of MCIM is close to 0 as long as the missingness of covariate is independent of the outcome, consistent with the conclusion in Appendices 1 and 2. We also conducted extensive numerical bias evaluations over a wide range of values typically encountered in epidemiologic settings for each parameter that determines the percent bias in MCIM. We found that the bias in MCIM was minimal in all but the most extreme cases. This result was further supported by empirical comparisons of the main results using the MCIM and multiple imputation in previously published studies.

Despite the ease of use, MCIM has generally been considered an unacceptably biased method for dealing with missing covariate data. This perception is largely based on two commonly cited studies published in 1990s [10, 11]. Vach and Belttner [11] performed the first quantitative investigation of the bias due to MCIM and concluded that "an important result of our empiric investigation is that creating an additional category for the missing values always yields biased results". However, this conclusion was drawn based on an assessment using extreme values for the parameters that determine the bias. For example, the authors used a relative risk of 0.36 for the outcome associated with covariate ($RR(C)$), and a relative risk of 9 for the relationship between the exposure and the covariate ($RR(E|C)$). Based on our numerical calculation, these values are indeed likely to produce biased results. However, these extreme values are rarely encountered in epidemiologic studies, and for them to occur simultaneously in a single study setting is extremely unlikely.

Similarly, in the simulation study by Greenland and Finkle [10], the missing covariate proportion was set at $50\%$, indicating that half of the participants had missing data, and the resulting values for $RR(E)$ produced by the MCIM ranged from 1.43 to 1.58 when the true $RR(E)$ was 2. Again, this relatively large bias is not surprising, given that, as shown in our **Table 3**, at $Pr(C_{miss})$=0.50, about $9\%$ of scenarios across the other parameters result in a $P_{bias}\%$ greater than $5\%$. Therefore, although it may first appear that our results are in conflict with the findings of these two previous studies, once the empirical values used in these investigations are considered, it is clear that while substantial bias can occur in rare, extreme cases, the bias is at most moderate in typical epidemiologic studies. This conclusion is consistent with the head-to-head comparisons of MCIM and multiple imputation in published studies from the Harvard cohorts [20-24].

While our derivation for the MCIM-related bias is based on the MCAR assumption, we also investigated under what circumstances the same formula (1) would apply when the missing covariate data mechanism is MAR. As demonstrated in Appendix 2, we found that the formula is valid in the case of MAR, as long as the missingness of the covariate is independent of the outcome. This would typically be the case in cohort studies and in nested case-control studies since $Y$ has not occurred at the start of the study, but may not be reasonable in population-based case-control studies. Unconditionally, the missingness of the covariate may depend on $Y$. But conditioning on other risk factors of $Y$, adjusted in the model, it can reasonably be assumed that the missingness of the covariate is independent of $Y$. It should be noted that other methods, including the maximum likelihood and multiple imputation approaches, are also based on the MAR assumption but allow the missingness of the covariate to depend on $Y$. Finally, although for ease of communication we considered only one additional risk factor for the outcome as a possible determinant of the MAR mechanism, it is a simple extension that an arbitrary number of $q$ additional covariates can be addressed by this work, simply by mapping them all into a single high or low risk indicator or through a propensity score.

In summary, through a comprehensive and systematic assessment using several approaches, we found no to minimal bias arising from the use of MCIM under a quite large range of circumstances that are typically encountered in epidemiologic studies. The continued use of MCIM is recommended unless the covariate is missing in an extreme proportion or acts as a strong confounder, with a relative risk for the outcome in relation to the confounder greater than 5 or with a very strong association between the exposure and confounder, both of which rarely occur in practice.

## Fund Support

This work was supported by a Mentored Research Scholar Grant in Applied and Clinical Research from the American Cancer Society (MRSG-17-220-01 – NEC to M.S.) and the National Institutes of Health (DPES025459 to D.S.).

## Data Availability Statement

The data that support the findings of this study are available from Harvard University, but restrictions apply to the availability of these data, which were used under license for the current study, and so are not publicly available. Data are however available from the authors upon reasonable request and with permission of Harvard University.

Table 1. Stratified 2×2 Tables for analysis using the missing covariate indicator method.

| | $C = 1$ | | $C = 0$ | | $C = missing$ | |
|---|---|---|---|---|---|---|
| | $E = 1$ | $E = 0$ | $E = 1$ | $E = 0$ | $E = 1$ | $E = 0$ |
| $Y = 1$ | $a_1$ | $b_1$ | $a_0$ | $b_0$ | $a_m$ | $b_m$ |
| $Y = 0$ | $c_1$ | $d_1$ | $c_0$ | $d_0$ | $c_m$ | $d_m$ |
| Total | $n_{11}$ | $n_{10}$ | $n_{01}$ | $n_{00}$ | $n_{m1}$ | $n_{m0}$ |

Table 2. The values of each parameter in numeric bias evaluation and simulation.

| | Parameter | Value |
|---|---|---|
| **Numeric Bias Evaluation** | $\Pr(C_{miss})$ | 0.005, **0.01, 0.05, 0.10**, 0.25, **0.50** |
| | Pr(E) | 0.01, 0.05, **0.10**, 0.25, **0.50**, 0.75 |
| | Pr(C) | 0.01, 0.05, **0.10**, 0.25, **0.50**, 0.75 |
| | RR(C) | 1/5, 1/3, 1/2, 1/1.5, 1/1.25, 1/1.15, **1**, 1.15, **1.25**, 1.5, 2, 3, 5 |
| | RR(E\|C) | 1/5, 1/3, 1/2, 1/1.5, 1/1.25, 1/1.15, **1**, 1.15, **1.25**, 1.5, 2, 3, 5 |
| **Simulation** | $\Pr(C_{miss})$ | **0.01, 0.05, 0.10, 0.50** |
| | Pr(E) | **0.10, 0.50** |
| | Pr(C) | **0.10, 0.50** |
| | RR(C) | 0.25, 0.75, **1**, **1.25**, 2.50 |
| | RR(E\|C) | 0.25, 0.75, **1**, **1.25**, 2.50 |
| | $exp\ (\alpha_1)$ | 1, 2, 5 |
| | $exp\ (\alpha_2)$ | 1, 2 |
| | Pr(Y) | 0.10, 0.50 |
| | RR(E) | 0.25, 0.75, 1, 1.25, 2.50 |

Overlapping parameter values are shown in bold. $\alpha_0$ is calculated to control the $\Pr(C_{miss})$ (see details in Appendix 4).

Table 3. Relative bias ($P_{bias}$%) in MCIM as a function of study parameters.

| Parameter value | Median | Percentile 25 | Percentile 75 | Percentage of $P_{bias}$%>10% | Percentage of $P_{bias}$%>5% |
|---|---|---|---|---|---|
| Overall | 0 | -0.09 | 0.09 | 1.13 | 2.61 |
| Pr(E) | | | | | |
| 0.01 | 0 | -0.07 | 0.07 | 0.81 | 2.07 |
| 0.05 | 0 | -0.08 | 0.07 | 0.81 | 2.07 |
| 0.1 | 0 | -0.08 | 0.08 | 0.84 | 2.15 |
| 0.25 | 0 | -0.09 | 0.09 | 1.28 | 2.73 |
| 0.5 | 0 | -0.11 | 0.11 | 1.43 | 3.15 |
| 0.75 | 0 | -0.15 | 0.14 | 1.90 | 3.99 |
| Pr(C) | | | | | |
| 0.01 | 0 | -0.01 | 0.01 | 0 | 0.07 |
| 0.05 | 0 | -0.05 | 0.05 | 0.31 | 0.87 |
| 0.1 | 0 | -0.10 | 0.10 | 0.82 | 1.87 |
| 0.25 | 0 | -0.20 | 0.21 | 1.71 | 3.88 |
| 0.5 | 0 | -0.24 | 0.26 | 2.17 | 4.99 |
| 0.75 | 0 | -0.20 | 0.21 | 1.73 | 3.90 |
| $Pr(C_{miss})$ | | | | | |
| 0.005 | 0 | -0.01 | 0.01 | 0 | 0 |
| 0.01 | 0 | -0.02 | 0.02 | 0 | 0 |
| 0.05 | 0 | -0.10 | 0.10 | 0 | 0.23 |
| 0.1 | 0 | -0.21 | 0.21 | 0.23 | 1.29 |
| 0.25 | 0 | -0.52 | 0.52 | 1.79 | 4.74 |
| 0.5 | 0 | -1.04 | 1.05 | 4.74 | 9.37 |
| RR(C) | | | | | |
| 1/5 | 0 | -0.24 | 0.28 | 3.02 | 5.81 |
| 1/3 | 0 | -0.18 | 0.22 | 2.09 | 4.65 |
| ½ | 0 | -0.13 | 0.16 | 1.01 | 2.79 |
| 1/1.5 | 0 | -0.08 | 0.09 | 0.19 | 1.28 |
| 1/1.25 | 0 | -0.05 | 0.05 | 0 | 0.23 |
| 1/1.15 | 0 | -0.03 | 0.03 | 0 | 0.04 |
| 1 | 0 | 0 | 0 | 0 | 0 |
| 1.15 | 0 | -0.04 | 0.03 | 0 | 0.04 |
| 1.25 | 0 | -0.06 | 0.05 | 0 | 0.23 |
| 1.5 | 0 | -0.11 | 0.10 | 0.19 | 1.40 |
| 2 | 0 | -0.21 | 0.19 | 1.16 | 3.37 |
| 3 | 0 | -0.34 | 0.32 | 2.64 | 5.74 |
| 5 | 0 | -0.54 | 0.51 | 4.34 | 8.29 |
| RR(E\|C) | | | | | |
| 1/5 | 0 | -0.30 | 0.23 | 2.60 | 5.13 |
| 1/3 | 0 | -0.24 | 0.18 | 1.83 | 4.12 |
| ½ | 0 | -0.17 | 0.13 | 1.09 | 2.75 |
| 1/1.5 | 0 | -0.12 | 0.09 | 0.48 | 1.65 |
| 1/1.25 | 0 | -0.07 | 0.05 | 0.18 | 0.64 |
| 1/1.15 | 0 | -0.05 | 0.03 | 0.07 | 0.28 |
| 1 | 0 | 0 | 0 | 0 | 0 |
| 1.15 | 0 | -0.04 | 0.05 | 0.07 | 0.36 |
| 1.25 | 0 | -0.06 | 0.08 | 0.21 | 0.82 |
| 1.5 | 0 | -0.10 | 0.14 | 0.52 | 2.12 |
| 2 | 0 | -0.17 | 0.24 | 1.57 | 4.05 |
| 3 | 0 | -0.30 | 0.40 | 2.94 | 6.32 |
| 5 | 0 | -0.48 | 0.67 | 4.82 | 8.97 |

Table 4. Summary table of simulation results of four methods, including no missing (NM), missing covariate indicator method (MCIM), multiple imputation (MI), and complete case (CC) with 2,000 replicates of each scenario.

| **Missing Mechanism** | **Method** | **Convergence Proportion** | **Mean Percent Relative Bias** | **Median Percent Relative Bias** | **Proportion \|Relative Bias\| $> 5\%$** | **Proportion \|Relative Bias\| $> 10\%$** | **Proportion \|Relative Bias\| $> 50\%$** |
|---|---|---|---|---|---|---|---|
| MCAR | NM | 0.86 | 0.40 | 0.09 | 0.15 | 0.07 | 0.00 |
| | MCIM | 0.85 | 0.41 | 0.08 | 0.24 | 0.13 | 0.01 |
| | MI | 0.69 | 0.45 | 0.08 | 0.17 | 0.08 | 0.00 |
| | CC | 0.86 | 1.32 | 0.11 | 0.17 | 0.09 | 0.01 |
| MAR $M$ independent of $Y$ | NM | 0.86 | 0.30 | 0.10 | 0.16 | 0.08 | 0.00 |
| | MCIM | 0.85 | 0.26 | 0.09 | 0.24 | 0.13 | 0.02 |
| | MI | 0.69 | 0.33 | 0.10 | 0.18 | 0.09 | 0.00 |
| | CC | 0.86 | 2.84 | 0.11 | 0.19 | 0.12 | 0.02 |
| MAR $M$ related to $Y$ | NM | 0.86 | 0.00 | 0.04 | 0.18 | 0.09 | 0.00 |
| | MCIM | 0.81 | -2.87 | -0.95 | 0.49 | 0.33 | 0.06 |
| | MI | 0.68 | 0.08 | 0.03 | 0.21 | 0.11 | 0.01 |
| | CC | 0.87 | 4.19 | -0.33 | 0.49 | 0.35 | 0.07 |

We include the scenarios that have at least one dataset with all four methods converged.

Table 5. Summary table of simulation results of asymptotic relative efficiency (ARE) of three methods, including missing covariate indicator method (MCIM), multiple imputation (MI), and complete case (CC) with respect to no missing (NM) with 2,000 datasets.

| Missing mechanism | Method | Mean | Median |
|---|---|---|---|
| MCAR | MCIM | 1.01 | 1.00 |
| | MI | 0.98 | 1.00 |
| | CC | 0.85 | 0.94 |
| MAR<br>$M$ independent of $Y$ | MCIM | 0.99 | 1.00 |
| | MI | 0.98 | 1.00 |
| | CC | 0.82 | 0.93 |
| MAR<br>$M$ related to $Y$ | MCIM | 1.01 | 1.00 |
| | MI | 0.98 | 1.00 |
| | CC | 0.75 | 0.86 |

We include the scenarios that have at least one dataset with all four methods converged. The ARE of MCIM with respect to NM is $ARE(MCIM, NM) = \frac{Var\left(\beta_1^{(NM)}\right)}{Var\left(\beta_1^{(MCIM)}\right)}$, where $\beta_1^{(NM)}$ and $\beta_1^{(MCIM)}$ are the exposure effect estimated by NM and MCIM, respectively. The ARE of MI and CC with respect to NM have similar form.

Table 6. Percentage of missingness (%) in the established breast cancer risk factors in the Nurses' Health Study (NHS) I and II*.

| | **NHS I (1976-2008)** | **NHS II (1989-2009)** |
|---|---|---|
| Age at 1st birth | 2.22 | 0 |
| Alcohol | 11.75 | 0.27 |
| History of benign breast disease | 0 | 0 |
| Family history breast cancer | 0 | 0 |
| Age at menarche | 0.90 | 0.33 |
| Age of menopause menopause | 9.37 | 6.57 |
| Physical Activity | 8.78 | 0.07 |
| Oral contraceptive use | 4.90 | 0.02 |
| Postmenopausal hormone use | 1.05 | 0.05 |
| Body mass index | 0.34 | 0.28 |
| Missing one or more of these covariates | 22.37 | 7.35 |

*Calculated as the proportion of person-years for which the variable had missing data among the total person-years.

Table 7. Comparison of the primary results in published studies that have used both missing covariate indicator method and multiple imputation.

| Publication Year (First Author) | Journal | Exposure of Interest | Outcome | Exposure Categories | Missing Covariate Indicator Method | Multiple Imputation |
|---|---|---|---|---|---|---|
| 2010 (Fung T) | Ann Intern Med | Low-carbohydrate diet score | All-cause mortality | 5th vs. 1st decile | 1.04 (0.96-1.12) | 1.06 (1.03-1.10) |
| 2012 (Joosten MM) | JAMA | Conventional cardiovascular risk factors | Peripheral artery disease | Ever smoking | 2.44 (1.98-3.00) | 2.43 (1.98-2.99) |
| | | | | Hypertension | 2.45 (2.01-2.98) | 2.47 (2.03-3.01) |
| | | | | Cholesterolemia | 1.42 (1.18-1.72) | 1.47 (1.22-1.77) |
| | | | | Diabetes | 2.45 (1.98-3.03) | 2.42 (1.96-2.90) |
| | | | | Per 1 unit increment in score | 2.10 (1.92-2.30) | 2.12 (1.94-2.32) |
| 2013 (Cahill LE) | Circulation | Breakfast eating | Coronary heart disease | Skipping Breakfast | 1.27 (1.06-1.53) | 1.29 (1.07-1.56) |
| | | | | Late night eating | 1.55 (1.05-2.29) | 1.53 (1.01-2.32) |
| | | | | Eating frequency | | |
| | | | | 1-2 times/day | 1.10 (0.91-1.31) | 1.17 (0.86-1.58) |
| | | | | 3 times/day | 1 (reference) | 1 (reference) |
| | | | | 4-5 times/day | 1.05 (0.94-1.18) | 1.05 (0.79-1.38) |
| | | | | 6+ times/day | 1.26 (0.90-1.77) | 1.21 (0.56-2.61) |
| 2013 (Pan A) | JAMA Intern Med | Change in red meat consumption | Type II diabetes | Decrease of >0.50 | 0.95 (0.84-1.07) | 0.98 (0.87-1.1) |
| | | | | Decrease of 0.15-0.50 | 0.98 (0.89-1.08) | 0.99 (0.9-1.095) |
| | | | | Change within ±0.14 | 1 (reference) | 1 (reference) |
| | | | | Increase of 0.15-0.50 | 1.10 (0.99-1.21) | 1.09 (0.98-1.21) |
| | | | | Increase of >0.50 | 1.22 (1.08-1.38) | 1.19 (1.06-1.35) |
| 2016 (Mu F) | Circ Cardiovasc Qual Outcomes | Endometriosis | Coronary heart disease | Endometriosis | 1.62 (1.39-1.89) | 1.63 (1.38-1.92) |

**Appendix 1. Derivation of the percentage of bias under MCAR**

**1. Parameter notations:**

$P_{bias}\%$: percentage of bias arising from missing covariate indicator method

Pr $(Y)$: cumulative incidence of outcome

$\Pr(E)$: prevalence of exposure

$\Pr(C)$: prevalence of covariate

$\Pr(C_{miss})$: proportion of missingness in covariate

$RR(E)$: expected relative risk of outcome associated with exposure

$RR_e$: estimated relative risk of outcome associated with exposure

$RR_{miss}$: crude estimate of the relative risk of outcome associated with exposure in the stratum with missing covariate (see **Table 1** in the main text)

$RR(C)$: expected relative risk of outcome associated with covariate

$RR(E|C)$: expected relative risk for the association between exposure and covariate

**2. Derivation of $P_{bias}\%$ under MCAR**

For this section, we assume that the missing covariate mechanism is missing completely at random (MCAR). The Appendix 2 demonstrates the sufficient conditions under which the same results would apply for missing at random (MAR).

The derivation is based on the complete data. Let $N_C$ be the size of complete data, i.e., $N_C = n_{00} + n_{01} + n_{10} + n_{11}$. Using the notation in **Table 1**,

$$RR_{miss} = \left(\frac{a_1 + a_0}{a_1 + a_0 + c_1 + c_0}\right) \Big/ \left(\frac{b_1 + b_0}{b_1 + b_0 + d_1 + d_0}\right) = \left(\frac{a_1 + a_0}{b_1 + b_0}\right)\left(\frac{b_1 + b_0 + d_1 + d_0}{a_1 + a_0 + c_1 + c_0}\right)$$

Given that $\frac{E(b_1+b_0+d_1+d_0)}{E(a_1+a_0+c_1+c_0)} = \frac{1-\Pr(E)}{\Pr(E)}$, next we derive $\frac{E(a_1+a_0)}{E(b_1+b_0)}$.

Based on **Table 1**, $a_1$ can be expressed as $E(a_1) = \Pr(Y = 1|E = 1, C = 1)E(n_{11})$.

Assume homogeneity of relative risks across the covariate $C$, that is,

$$RR(E) = \frac{\Pr\,(Y=1|E=1, C=1)}{\Pr\,(Y=1|E=0, C=1)} = \frac{\Pr\,(Y=1|E=1, C=0)}{\Pr\,(Y=1|E=0, C=0)}.$$

Identically,

$$RR(C) = \frac{\Pr\,(Y=1|E=1, C=1)}{\Pr\,(Y=1|E=1, C=0)} = \frac{\Pr\,(Y=1|E=0, C=1)}{\Pr\,(Y=1|E=0, C=0)}.$$

Based on **Table 1**, we have

$E(a_1) = \Pr(Y=1|E=1, C=1)E(n_{11}) = \Pr(Y=1|E=1, C=1)\Pr(C=1|E=1)\Pr\,(E=1)\,N_C$.

Similarly,

$E(a_0) = \Pr(Y=1|E=1, C=0)\Pr(C=0|E=1)\Pr\,(E=1)\,N_C$,

$E(b_1) = \Pr(Y=1|E=0, C=1)\Pr(C=1|E=0)\Pr\,(E=0)N_C$, and

$E(b_0) = \Pr(Y=1|E=0, C=0)\Pr(C=0|E=0)\Pr(E=0)\,N_C$.

By the law of total probability rule,

$\Pr(E=1) = \Pr(E=1|C=1)\Pr(C) + \Pr(E=1|C=0)\,[1-\Pr(C)]$

$= RR(E|C)\Pr(E=1|C=0)\Pr(C) + \Pr(E=1|C=0)\,[1-\Pr(C)]$

$= \Pr(E=1|C=0)\,[RR(E|C)\Pr(C) + 1 - \Pr(C)]$

Therefore, $\Pr(E=1|C=0) = \frac{\Pr\,(E)}{RR(E|C)\Pr(C)+1-\Pr(C)}$ and $\Pr(E=1|C=1) = \frac{\mathrm{RR(E|C)Pr}\,(E)}{RR(E|C)\Pr(C)+1-\Pr(C)}$.

Then,

$$\Pr(C=1|E=1) = \frac{\Pr(\mathrm{E}=1|\mathrm{C}=1)\Pr\,(\mathrm{C})}{\Pr(E)} = \frac{\mathrm{RR(E|C)Pr}\,(C)}{RR(E|C)\Pr(C)+1-\Pr(C)}. \qquad \text{(A1)}$$

By the law of total probability rule again,

$\Pr(C=1) = \Pr(C=1|E=1)\Pr(E) + \Pr(C=1|E=0)\,[1-\Pr(E)]$. Then,

$$\Pr(C=1|E=0)=\frac{\Pr(C)-\Pr(C=1|E=1)\Pr(E)}{1-\Pr(E)} \quad \text{(A2)}$$

Therefore,

$$\frac{E(a_1+a_0)}{E(b_1+b_0)} = \frac{\Pr(Y=1|E=1,C=1)\Pr(C=1|E=1)+\Pr(Y=1|E=1,C=0)\Pr(C=0|E=1)}{\Pr(Y=1|E=0,C=1)\Pr(C=1|E=0)+\Pr(Y=1|E=0,C=0)\Pr(C=0|E=0)} * \frac{\Pr(E)}{1-\Pr(E)}.$$

Then, we have

$$RR_{miss} \xrightarrow{p} \frac{\Pr(Y=1|E=1,C=1)\Pr(C=1|E=1)+\Pr(Y=1|E=1,C=0)\Pr(C=0|E=1)}{\Pr(Y=1|E=0,C=1)\Pr(C=1|E=0)+\Pr(Y=1|E=0,C=0)\Pr(C=0|E=0)}$$
$$= RR(E)\frac{RR(C)\Pr(C=1|E=1)+1-\Pr(C=1|E=1)}{\mathrm{RR(C)}\Pr(\mathrm{C}=1|\mathrm{E}=0)+1-\Pr(\mathrm{C}=1|\mathrm{E}=0)}.$$

Given that $RR_e=[1-\Pr(C_{miss})]RR(E)+\Pr(C_{miss})\,RR_{miss}$

$$P_{bias}\% = \frac{RR_e-RR(E)}{RR(E)}\times 100 = \frac{[1-\Pr(C_{miss})]RR(E)+\Pr(C_{miss})\,RR_{miss}-RR(E)}{RR(E)}\times 100$$
$$= \Pr(C_{miss})\left(\frac{RR_{miss}}{RR(E)}-1\right)\times 100.$$

Then, we have,

$$P_{bias}\% = \Pr(C_{miss})\left(\frac{[RR(C)-1]\,[\Pr(C=1|E=1)-\Pr(C=1|E=0)]}{\mathrm{RR(C)}\Pr(\mathrm{C}=1|\mathrm{E}=0)+1-\Pr(\mathrm{C}=1|\mathrm{E}=0)}\right)\times 100. \quad \text{(A3)}$$

Substituting (A1) and (A2) into (A3), after some simple algebras, we obtain,

$$P_{bias}\% = \Pr(C_{miss})\frac{\Pr(C)\,[1-\Pr(C)]\,[RR(C)-1][RR(E|C)-1]}{[1-\Pr(E)][1-\Pr(C)+\Pr(C)\,RR(C)RR(E|C)]-\Pr(C)[1-\Pr(C)][RR(C)-1][RR(E|C)-1]}\times 100.$$

Notice that the dependence of this expression on $\Pr(Y)$ and on $RR(E)$ is eliminated.

## 3. Some special cases

1) When $\Pr(C_{miss}) \to 0$, then $P_{bias}\% \to 0$.
2) When $\Pr(C) \to 0$ or $\Pr(C) \to 1$, then $P_{bias}\% \to 0$.
3) When $RR(C) = 1$ (i.e., covariate has no effect on outcome), then $P_{bias}\% = 0$.
4) When $RR(E|C) = 1$ (i.e., covariate is not associated with exposure), then $P_{bias}\% = 0$.

For Cases 2, 3 and 4, the covariate is not a confounder of the exposure-outcome relationship, and hence there is no bias.

## 4. Restriction on the parameters

The values of each parameter considered in this paper are given in **Table 2**. However, some combinations of the values are invalid, in the sense that they produce probabilities outside the range of [0,1]. To see this, we calculated the following probabilities,

$$\Pr(E = 1|C = 0) = \frac{\Pr(E)}{RR(E|C)\Pr(C)+1-\Pr(C)},\ \Pr(E = 1|C = 1) = \frac{RR(E|C)\Pr(E)}{RR(E|C)\Pr(C)+1-\Pr(C)},$$

$$\Pr(C = 1|E = 1) = \frac{\mathrm{RR(E|C)}\Pr(C)}{RR(E|C)\Pr(C)+1-\Pr(C)}, \text{ and } \Pr(C = 1|E = 0) = \frac{\Pr(C)-\Pr(C=1|E=1)\Pr(E)}{1-\Pr(E)}.$$

Note that $\Pr(C = 1|E = 1)$ will always be between 0 and 1, but the others are not so restricted. We excluded from the numerical evaluation of $P_{bias}\%$ the sets of parameter values that produced $\Pr(C = 1|E = 0)$, $\Pr(E = 1|C = 0)$ or $\Pr(E = 1|C = 1)$ outside of 0 and 1. For example, when $\Pr(C) = 0.5$, $\Pr(E) = 0.75$, and $RR(E|C) = 0.5$, using the above formulae, we have $\Pr(C = 1|E = 0) = 1.5$, $\Pr(E = 1|C = 0) = 1.25$, and $\Pr(E = 1|C = 1) = 0.25$. Thus, this combination of $\Pr(C) = 0.5$, $\Pr(E) = 0.75$, and $RR(E|C) = 0.5$ is not valid, and was excluded from the evaluation of $P_{bias}\%$.

## Appendix 2. Sufficient conditions for the results in Appendix 1 to apply when the missing mechanism is MAR

In Appendix 1, we assumed that the covariate is missing completely at random (MCAR). Under MCAR, the RR estimated from the complete observations is the true RR, and the crude RR in the stratum with missing covariate is equal to the crude RR in the strata with complete observations.

In this appendix, we consider the case of missing at random (MAR), and investigate the condition for the results of Appendix 1 to apply.

Variables are defined as in the rest of this manuscript. Let $N$ be the total sample size of the study. Under MAR, the missingness of covariate $C$ may depend on $E$ and $Y$, but is independent of $C$. Let $f_{ye} = \Pr(M = 0|Y = y, E = e)$ be the probability of no missing $C$, where $M = 1$ if the $C$ is missing and 0 otherwise. When $C = c$,

$$E(a_c) = N \Pr(C = c) \Pr(E = 1|C = c) \Pr(Y = 1|E = 1, C = c) f_{11}.$$

Similarly,

$$E(b_c) = N \Pr(C = c) \Pr(E = 0|C = c) \Pr(Y = 1|E = 0, C = c) f_{10},$$

$$E(c_c) = N \Pr(C = c) \Pr(E = 1|C = c) \Pr(Y = 0|E = 1, C = c) f_{01},$$

$$E(d_c) = N \Pr(C = c) \Pr(E = 0|C = c) \Pr(Y = 0|E = 0, C = c) f_{00}.$$

Assume again the RR is equal across the strata $C = 1$ and $C = 0$. In the stratum $C = c$, when the sample size approaches infinity, the observed RR

$$RR_c = \frac{a_c}{a_c + c_c} \Big/ \frac{b_c}{b_c + d_c} \xrightarrow{p}$$

$$\frac{\Pr(Y = 1|E = 1, C = c)}{\Pr(Y = 1|E = 0, C = c)} \times \frac{f_{11}}{f_{10}} \times \frac{\Pr(Y = 1|E = 0, C = c)(f_{10} - f_{00}) + f_{00}}{\Pr(Y = 1|E = 1, C = c)(f_{11} - f_{01}) + f_{01}}$$

$$= RR(E) \times \frac{f_{11}}{f_{10}} \times \frac{\Pr(Y = 1|E = 0, C = c)(f_{10} - f_{00}) + f_{00}}{\Pr(Y = 1|E = 1, C = c)(f_{11} - f_{01}) + f_{01}}.$$

So when $f_{10} = f_{00}$ and $f_{11} = f_{01}$, the observed RR converges to the true RR, $RR(E)$. In addition, the observed RR converges to the true RR when $\frac{f_{11}}{f_{10}} \times \frac{\Pr(Y = 1|E = 0, C = c)(f_{10} - f_{00}) + f_{00}}{\Pr(Y = 1|E = 1, C = c)(f_{11} - f_{01}) + f_{01}} = 1$. Thus, when $f_{10} = f_{00}$ and $f_{11} = f_{01}$, that is, when the probability of missingness of the covariate, $C$, is independent of the outcome, $Y$, but may depend on the exposure, $E$, the observed RR converges to the true RR.

We next investigate the conditions under which the crude RR in the missing stratum is equal to the crude RR in the complete data strata. In the missing stratum, we have

$$\begin{aligned} E(a_m) = N[&\Pr(\mathrm{C}=1)\Pr(E=1|C=1)\Pr(Y=1|E=1,C=1) \\ &+ \Pr(\mathrm{C}=0)\Pr(E=1|C=0)\Pr(Y=1|E=1,C=0)](1-f_{11}) \\ &= E(a_1+a_0)\times\frac{1-f_{11}}{f_{11}}. \end{aligned}$$

Similarly,

$$E(b_m) = E(b_1+b_0)\times\frac{1-f_{10}}{f_{10}}$$

$$E(c_m) = E(c_1+c_0)\times\frac{1-f_{01}}{f_{01}}$$

$$E(d_m) = E(d_1+d_0)\times\frac{1-f_{00}}{f_{00}}$$

The estimated crude RR in the missing stratum is $\frac{a_m}{a_m+c_m}\Big/\frac{b_m}{b_m+d_m}$, and the estimated crude RR in the complete observation strata is $\frac{a_1+a_0}{a_1+a_0+c_1+c_0}\Big/\frac{b_1+b_0}{b_1+b_0+d_1+d_0}$.

When $f_{10} = f_{00}$ and $f_{11} = f_{01}$, we have

$$\frac{E(a_m)}{E(a_m+c_m)}\Bigg/\frac{E(b_m)}{E(b_m+d_m)} = \frac{E(a_1+a_0)}{E(a_1+a_0+c_1+c_0)}\Bigg/\frac{E(b_1+b_0)}{E(b_1+b_0+d_1+d_0)}.$$

Thus, under the conditions that $f_{10} = f_{00}$ and $f_{11} = f_{01}$, two crude RRs asymptotically coincide.

Again, the conditions $f_{10} = f_{00}$ and $f_{11} = f_{01}$ imply that the probability of missingness of the covariate $C$ is independent of the outcome $Y$, but may depend on the exposure $E$, and the results of Appendix 1 continue to hold.

**Appendix 3. Sufficient conditions for the results in Appendix 2 to apply for odds ratio (OR) under MAR**

In appendix 2, we investigated the conditions that the observed RR converges to the true RR when the missing mechanism is MAR.

For rare diseases, the OR closely approximates the RR. In this appendix, we explore the conditions under which the observed OR converges to the true OR when the missing mechanism is MAR.

Under MAR, the missingness of covariate $C$ may depend on $E$ and $Y$, but is assumed to be independent of $C$. Additionally, we assume that the OR is equal across the strata $C = 1$ and $C = 0$, i.e.,

$$OR(E) = \frac{\Pr(Y=1|E=1,C=1)}{1-\Pr(Y=1|E=1,C=1)} \Big/ \frac{\Pr(Y=1|E=0,C=1)}{1-\Pr(Y=1|E=0,C=1)} = \frac{\Pr(Y=1|E=1,C=0)}{1-\Pr(Y=1|E=1,C=0)} \Big/ \frac{\Pr(Y=1|E=0,C=0)}{1-\Pr(Y=1|E=0,C=0)}.$$

In the stratum $C = c$, when the sample size approaches infinity, the observed OR

$$OR_c = \frac{a_c}{c_c} \Big/ \frac{b_c}{d_c} \xrightarrow{p} \frac{\Pr(Y=1|E=1,C=c)}{1-\Pr(Y=1|E=1,C=c)} \times \frac{f_{11}}{f_{10}} \times \frac{1-\Pr(Y=1|E=0,C=c)}{\Pr(Y=1|E=0,C=c)} \times \frac{f_{00}}{f_{01}}$$
$$= OR(E) \times \frac{f_{11}f_{00}}{f_{10}f_{01}}.$$

Thus, when $\frac{f_{11}f_{00}}{f_{10}f_{01}} = 1$, the observed OR converges to the true OR. Several specific cases satisfy this condition:

1) **Independence of missingness in covariate and outcome:** when $f_{10} = f_{00}$ and $f_{11} = f_{01}$, that is, when the probability of missingness of the covariate $C$ is independent of the outcome $Y$, but may depend on the exposure $E$.
2) **Independence of missingness in covariate and exposure:** when $f_{10} = f_{11}$ and $f_{01} = f_{00}$, that is, when the probability of missingness of the covariate $C$ is independent of the exposure $E$, but may depend on the outcome $Y$.
3) **Generalized linear model for missingness in covariate:** when $f_{ye} = \exp(\gamma_0 + \gamma_1 y + \gamma_2 e)$, where $\gamma_0$, $\gamma_1$, and $\gamma_2$, are parameters for intercept, $Y$, and $E$, respectively.

We next investigate the conditions under which the crude OR in the missing stratum is equal to the crude OR in the complete data strata.

The estimated crude OR in the missing stratum is $\frac{a_m}{c_m} \Big/ \frac{b_m}{d_m}$, and the estimated crude OR in the complete observation strata is $\frac{a_1+a_0}{c_1+c_0} \Big/ \frac{b_1+b_0}{d_1+d_0}$.

Then

$$\frac{E(a_m)}{E(a_m+c_m)} \Big/ \frac{E(b_m)}{E(b_m+d_m)} = \left[\frac{E(a_1+a_0)}{E(a_1+a_0+c_1+c_0)} \Big/ \frac{E(b_1+b_0)}{E(b_1+b_0+d_1+d_0)}\right] \frac{f_{10}f_{01}(1-f_{11})(1-f_{00})}{f_{11}f_{00}(1-f_{10})(1-f_{01})}.$$

Thus, under the conditions that $\frac{f_{10}f_{01}(1-f_{11})(1-f_{00})}{f_{11}f_{00}(1-f_{10})(1-f_{01})} = 1$, two crude ORs asymptotically coincide.

Again, several specific cases satisfy this condition:

1) **Independence of missingness in covariate and outcome:** when $f_{10} = f_{00}$ and $f_{11} = f_{01}$, that is, when the probability of missingness of the covariate $C$ is independent of the outcome $Y$, but may depend on the exposure $E$.
2) **Independence of missingness in covariate and exposure:** when $f_{10} = f_{11}$ and $f_{01} = f_{00}$, that is, when the probability of missingness of the covariate $C$ is independent of the exposure $E$, but may depend on the outcome $Y$.

Thus, for rare disease, the results in Appendix 1 remain valid when the missingness of the covariate $C$ is independent of either the outcome $Y$ or the exposure $E$.

## Appendix 4. Simulation settings

We consider all parameter combinations in **Table 2**, excluding scenarios with invalid probability of $\Pr(E|C)$ or $\Pr(Y|E,C)$, i.e., $\Pr(E|C) < 0, \Pr(E|C) > 1, \Pr(Y|E,C) < 0$, or $\Pr(Y|E,C) > 1$, and scenarios with $\Pr(E) = \Pr(C) = \Pr(Y) = 0.1$ to ensure $a_1$ in **Table 1** is sufficiently large. For each scenario, we simulated the datasets 2,000 times. We simulated the dataset $(Y_i, E_i, C_i, M_i)$ for $i = 1, \cdots, 2{,}000$ with two missing data mechanisms: MCAR and MAR. For MCAR, we studied a total of 4,704 scenarios. For MAR, we studied a total of 28,224 scenarios.

We first simulated the covariate $C_i \sim \text{Bernoulli}(\Pr(C))$. Next, we simulated the exposure $E_i|C_i \sim \text{Bernoulli}(\Pr(E|C_i))$, where $\Pr(E|C_i) = \frac{\text{RR(E|C)}^{C_i}\Pr(E)}{RR(E|C)\Pr(C)+1-\Pr(C)}$. Next, we simulated the outcome $Y_i|E_i, C_i \sim \text{Bernoulli}(\Pr(Y|E_i, C_i))$, where

$$\Pr(Y|E_i, C_i) = RR(E)^{E_i}RR(C)^{C_i}\,Pr(Y)\,/[(1 - Pr\,(E = 1|C = 0))(1 - Pr\,(C)) + RR(E)\,Pr\,(E = 1|C = 0)(1 - Pr\,(C)) + RR(C)(1 - Pr\,(E = 1|C = 1))\,Pr\,(C) + RR(E)RR(C)\,Pr\,(E = 1|C = 1)\,Pr\,(C)].$$

Lastly, we simulated the missing indicator $M_i$. For MCAR, we simulated the $M_i$~Bernoulli($\Pr(C_{miss})$). For MAR, we simulated the $M_i$ as $logit\big(P(M_i = 1|E_i, Y_i)\big) = \alpha_0 + \alpha_1 E_i + \alpha_2 Y_i$, where $\alpha_1$ and $\alpha_2$ are given in **Table 2**, $\alpha_0$ is calculated to control the $\Pr(C_{miss})$ based on the following equation:

$$\Pr(C_{miss}) = \frac{\exp(\alpha_0+\alpha_1+\alpha_2)}{1+\exp(\alpha_0+\alpha_1+\alpha_2)}P(E = 1, Y = 1) + \frac{\exp(\alpha_0+\alpha_1)}{1+\exp(\alpha_0+\alpha_1)}P(E = 1, Y = 0) + \frac{\exp(\alpha_0+\alpha_2)}{1+\exp(\alpha_0+\alpha_2)}P(E = 0, Y = 1) + \frac{\exp(\alpha_0)}{1+\exp(\alpha_0)}P(E = 0, Y = 0),$$

with

$$P(E = 1, Y = 1) = P(Y = 1|E = 1, C = 1)P(E = 1|C = 1)\,Pr(C) + P(Y = 1|E = 1, C = 0)P(E = 1|C = 0)(1 - Pr(C)),$$

$$P(E = 1, Y = 0) = \big(1 - P(Y = 1|E = 1, C = 1)\big)P(E = 1|C = 1)\,\Pr(C) + \big(1 - P(Y = 1|E = 1, C = 0)\big)P(E = 1|C = 0)(1 - Pr(C)),$$

$$P(E = 0, Y = 1) = P(Y = 1|E = 0, C = 1)\big(1 - P(E = 1|C = 1)\big)\,Pr(C) + P(Y = 1|E = 0, C = 0)(1 - P(E = 1|C = 0))(1 - Pr(C)),$$

and $P(E = 0, Y = 0) = 1 - P(E = 1, Y = 1) - P(E = 1, Y = 0) - P(E = 0, Y = 1)$.